\documentclass[traditabstract,printer]{aa}
\usepackage{txfonts}
\usepackage{graphicx}
\usepackage{color}

\newcommand\phs{\phantom{$-$}}

\begin{document}

\title{Neutron-capture element deficiency of the Hercules dwarf spheroidal galaxy
\thanks{Based on observations carried out at the European Southern Observatory under proposal number 083.D-0688(A).}
}

\author{Andreas Koch\inst{1}  
 \and Sofia Feltzing\inst{2}
 \and Daniel Ad\'en\inst{2} 
\and Francesca Matteucci\inst{3} 
}
\authorrunning{A. Koch  et al.}
\titlerunning{Barium abundances in the Hercules dwarf spheroidal}
\offprints{A. Koch\\ \email{akoch@lsw.uni-heidelberg.de}}

\institute{
Zentrum f\"ur Astronomie der Universit\"at Heidelberg, Landessternwarte, K\"onigstuhl 12, 69117 Heidelberg, Germany
\and
Lund Observatory, Box 43, SE-22100 Lund, Sweden
\and
Dipartimento di Astronomia, Universit\'a di Trieste, via G. B. Tiepolo 11, 34143 Trieste, Italy
}
\date{}

\abstract {We present an assessment of  the barium abundance ratios for red giant member stars in the faint Hercules dwarf spheroidal (dSph) galaxy. 
Our results are drawn from intermediate-resolution FLAMES/GIRAFFE spectra around the Ba~II 6141.71 \AA~absorption line at  low signal-to-noise ratios. For
three brighter stars we were able to gain estimates from direct equivalent-width measurements, while for the remaining eight stars only upper limits could be obtained. 
These results are investigated in a statistical manner and indicate 
very low Ba abundances of log\,$\varepsilon$(Ba)$\la$ 0.7 dex (3$\sigma$).
We discuss various possible systematic biasses, first and foremost, a blend with the Fe~I 6141.73 \AA-line, but most of those would only lead to even lower abundances. 
A better match with metal-poor halo and dSph stars can only be reached by including a large uncertainty in the continuum placement. 
This contrasts with the high dispersions in iron and calcium (in excess of 1 dex) in this galaxy.   
While the latter spreads are typical of the very low luminosity, dark-matter dominated dSphs, a high level of depletion in heavy elements suggests 
that chemical enrichment in Hercules was governed by very massive stars, coupled with a very low star formation efficiency.
While very low abundances of some heavy elements are also found in individual stars of other dwarf galaxies, 
this is the first time that a very low Ba abundance is found {\em within an entire} dSph over a broad metallicity range. 
}

\maketitle 
%
%
%
%
%
%
\section{Introduction}
Dwarf spheroidal (dSph) galaxies are low-luminosity systems with total
luminosities often not exceeding those of globular clusters (e.g.,
Mateo 1998). Recently, even fainter systems have been found (Zucker et
al. 2006a,b; Belokurov et al. 2006; Belokurov et al. 2007; Belokurov
et al. 2009). These {\em ultra-}faint dSphs (UFDs) are exteremely
faint and have very high mass-to-light ratios (e.g., Koch 2009; Simon
et al. 2011).  The UFDs are predominantly old and metal-poor systems,
on average, more metal poor than their luminous counterparts,
which extends the metallicity-luminosity relation toward even fainter
magnitudes (Kirby et al. 2008).  While long evasive (Fulbright et
al. 2004)  very metal-poor stars below an [Fe/H] of $-3$ dex have now
been uncovered in the more luminous dSphs as well as the UFDs
(Kirby et al. 2008), with the currently most iron-poor star lying at
[Fe/H]=$-3.96$ dex (Tafelmeyer et al. 2010).

The UFDs were probably the formation sites of some of the first
stars in the Universe (e.g., Salvadori \& Ferrara; Bovill \& Ricotti
2009; Gao et al. 2010; Norris et al. 2010a).  Given their low stellar
masses, the UFDs probably only hosted a few massive supernovae (SNe) so that they
are the ideal environments to look for the imprints of {\em
  individual} explosions in the next generation of stars that formed
out of the earliest SN material (e.g., Iwamoto et al. 2005; Lai et al
2008).

An important clue to the formation and evolution of the Galactic halo
lies in the detection of anomalies in elemental abundances in a few halo
field stars (e.g., Aoki et al. 2002; Ivans et al. 2003; Cohen et
al. 2007) such as selective enhancements and/or depletions in the
$\alpha$- or neutron-capture elements.  The discovery of chemical
oddballs in a few dSph satellites (e.g., Fulbright et al. 2004; Koch et al. 2008a; Feltzing et
al. 2009) appears to indicate that accreted and disrupted UFDs were in
fact important early donors to the metal-poor component of the
Galactic halo.  By studying the elemental abundances in more UFDs we
can probe their similarity to the anomalous halo objects and assess
how common such stars are in the dSphs, and investigate the
predominant sources for their anomalous abundance patterns.

Koch et al. (2008, hereafter K08) studied two red giant branch (RGB) stars in the Hercules dSph galaxy 
in detail and found an ``extraordinary level of depletion'' in heavy
elements. The level of depletion is reminiscent of what has previously
been found for the metal-poor red giant star Dra~119 in the Draco dSph
galaxy (Fulbright et al. 2004), albeit at a significantly higher
metallicity.  Similar abundance patterns were also measured more recently in
three stars of the ultra-faint dSph galaxies Bo\"otes~I (Feltzing et
al. 2009) and Leo~IV (Simon et al. 2010).  Both galaxies have even
lower baryonic masses than Hercules. This means that we can expect to see
the signatures of individual SNe, manifested in peculiar abundance
patterns for the elements as well as a large star-to-star scatter in
the abundance ratios (see also Carigi \& Hernandez 2008; Koch et
al. 2008a; Marcolini et al. 2008). K08 argued that the peculiar
abundance patterns measured in the Hercules dSph galaxy, such as abnormally high Mg/Ca and
Co/Cr ratios, were caused by an incomplete sampling of the initial
mass function so that most likely only 1--3 massive ($\sim$35 M$_{\odot}$)
SNe of type II enriched the galaxy's gas.

In this work we focus on abundance measurements of the neutron-capture
element barium ($Z=56$) in the Hercules dSph galaxy (Belokurov et
al. 2007). At $M_V=-6.6$ mag, Hercules is one of the brighter UFDs.  Hercules is
a typical, metal-poor, old stellar population ($>$12 Gyr) that has
experienced little to no star formation in the past 12 Gyr (Sand et
al. 2009; Musella et al. 2012; but cf. Brown et al. 2012).
Narrow-band photometry (Ad\'en et al. 2009a) and medium- to
high-resolution spectroscopy (K08; Ad\'en et al. 2009a;
Ad\'en et al. 2011 [hereafter A11]) have uncovered a broad range in
``metallicity'' in this galaxy with a full range in [Fe/H] from $-3.2$
to $-2$ dex, indicative of the presence of significant amounts of dark matter (Koch et al. 2012a)\footnote{In fact, its mass-to-light ratio is very high ($\sim$300), implying a stellar mass of 
only a few times 10$^4$ M$_{\odot}$  (Martin et al. 2008; see also Strigari et al. 2008; Ad\'en et al. 2009b).}. 

This paper is organized as follows: In \textsection2 we briefly
recapitulate details on the stellar sample and the data set that we
base our study on.  In \textsection3 we describe our estimates of the
Ba abundance in the Hercules stars including a comprehensive discussion of
possible biases of our measurements. The results are presented in
\textsection4 and are compared with the literature in \textsection5 before we discuss these in the light of Hercules'
chemical enrichment history in \textsection6.
\section{Data and sample parameters}
Ad\'en et al. (2011) obtained spectra of 20 RGB stars selected from the comprehensive list of confirmed member stars of Ad\'en et al. (2009a) that was compiled  
using the HR13 grating of the Fibre Large Array Multi Element Spectrograph (FLAMES;  Pasquini et al. 2002) at the Very Large Telescope (VLT). 
This setting provides a wavelength coverage of 6100--6400 \AA~at a resolving power of R=20000. We refer to A11 for details on the 
observation and reduction strategy for this data set. Typical signal-to-noise (S/N) ratios 
for the entire sample in the magnitude range of V=18.7--20.3 mag are
reported as 8--35 per pixel, while the abundance analysis of A11 was
restricted to the 11 brighter targets (V$>$20 mag) with S/N ratios in
excess of 12.

In summary, A11 derived the atmospheric parameters of their RGB
targets from a combination of St\"omgren photometry and excitation
equilibrium ($T_{\rm eff}$), old, metal-poor isochrones (log\,$g$),
and using empirical relations between gravity and microturbulence ($\xi$). 
 The resulting element abundance ratios for iron and
calcium were then derived by means of an equivalent-width (EW)
analysis, complemented by the fitting of synthetic spectra for the
lower-S/N cases.
In the following, we will proceed by adopting the stellar parameters as well as the Fe and Ca abundances of A11. 
\section{Barium abundance measurements}
Throughout our analysis we used the Kurucz atmosphere
grid\footnote{{\tt http://kurucz.harvard.edu}}, operating in local
thermodynamic equilibrium (LTE), without convective overshoot, and
using the $\alpha$-enhanced opacity distributions AODFNEW (Castelli \&
Kurucz 2003)\footnote{{\tt http://wwwuser.oat.ts.astro.it/castelli}}.
The latter is justified by the findings of K08 that the abundances of
important electron donors (Mg, O) are highly elevated to above 0.8 dex
(Sect.~3.6.3), while the other $\alpha$-elements cover a broad range,
with, e.g., [Ca/Fe] spanning $-$0.3 to 0.3 dex (K08; A11).
Stellar abundances for the absorption lines in question were then
computed using the {\em abfind} and {\em synth} drivers of the 2010
version of the synthesis program MOOG (Sneden 1973).

We note that A11 used spherical MARCS model atmospheres (Gustafsson et
al. 2008) in their derivation of stellar parameters and [Fe/H] and
[Ca/H], while the present work uses Kurucz model atmospheres (in
following, e.g., K08). Heiter \& Eriksson (2006) showed that
significant differences can occur in derived elemental abundances for
evolved giants, such as those in the Hercules dSph galaxy. However, we
note that the differences in derived abundances, e.g., for Fe\,{\sc i}
and Fe\,{\sc ii} are small when the EWs are small (less than 50
m\AA). Because we are exclusively studying metal-poor stars, where the EWs 
are small, the effect should be much weaker than the uncertainties caused,
e.g., by continuum setting and S/N. Heiter \& Eriksson (2006) did not
include lines from Ba\,{\sc ii} in their study. That our results are
robust with respect to the usage of plane parallel or spherical model
atmospheres  is also corroborated by the investigation {in} Koch \&
McWilliam (2008). These authors showed that switching from the spherical MARCS
model atmospheres to the plane-parallel model atmospheres in the
Kurucz grid does not incur systematic changes in excess of 0.03 dex
(for Fe\,{\sc i} and Fe\,{\sc ii} lines) and we do not
expect any systematic effects on the limits that we will place on the
[Ba/Fe] abundance ratios in the following.

Finally, in all steps in the analysis we account for
hyperfine splitting of the Ba line. We used the same data for the
hyperfine structure as in McWilliam (1998). 
We note that
the effect of the hyperfine structure on the derived Ba abundance is
consistent with zero.
This is in line with laboratory work, which shows that the shifts of the main
isotopes are small and hence the line
is essentially Gaussian  in spectra of moderate resolution (Karlsson, priv. comm. and Karlsson \&
Litz\'en 1999;  Bensby et al. 2005).  We adopt the solar abundance
scale of Asplund et al. (2009), with log\,$\varepsilon_{\odot}$(Ba)=2.17.
\subsection{Equivalent-width measurements of the Ba line}
The S/N ratios in the three brightest stars, at $S/N=25-35$, allowed
us to attempt a direct measurement of the EW of the Ba\,{\sc ii} line
at 6141\,{\AA} -- the only neutron-capture absorption feature that
falls within the covered wavelength range and that would still be strong
enough to be measured in metal-poor stars such as the red giant stars
in the Hercules dSph galaxy. In practice, the EWs were measured by
adopting a Gaussian line profile using IRAF's {\em splot}
task. 
Measurement uncertainties on the EWs were
assigned based on the S/N and a generous allowance for the 
placement of the continuum.
Figure\,1 shows the respective region in the spectra of the three
giant stars. 
Table~1 lists the resulting EWs.
\begin{figure}[htb]
\begin{center}
\includegraphics[angle=0,width=1\hsize]{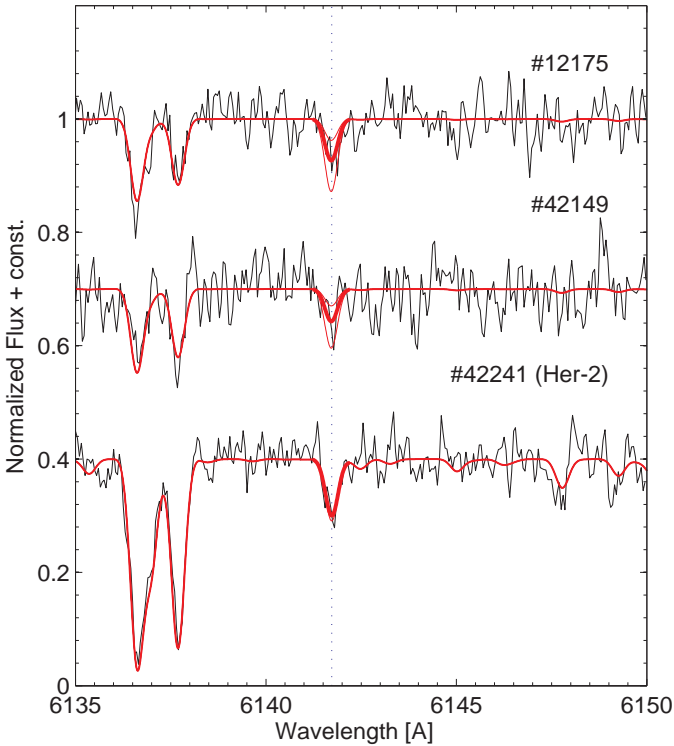}
\end{center}
\caption{Spectra of the three brightest stars for which an EW could
  be measured for the Ba~II 6141 \AA~line (indicated with a dotted 
  vertical line). The strongest features in this wavelength range are
  the two Fe\,{\sc i} lines to the left in the spectra (6136--6138
  \AA).  Shown in red are synthetic spectra with the best-fit
  Ba-abundance (thick lines) and bracketing
  log\,$\varepsilon$(Ba)$\pm$0.4 dex (thin lines).}
\end{figure}
\begin{table*}[htb]
\caption{Barium abundance  and upper limits derived as discussed in Sect.\,3.2.}             
\centering          
\begin{tabular}{ccccccccc}     
\hline\hline       
& & & EW  & \multicolumn{5}{c}{log\,$\varepsilon$(Ba)\tablefootmark{b}}\\  
\cline{5-9}
 \raisebox{1.5ex}[-1.5ex]{ID\tablefootmark{a}} &  
 \raisebox{1.5ex}[-1.5ex]{[Fe/H]\tablefootmark{a}} &  
 \raisebox{1.5ex}[-1.5ex]{[Ca/H]\tablefootmark{a}} & [m\AA] & EW &  Synth & 1$\sigma$ limit\tablefootmark{d} & 2$\sigma$ limit\tablefootmark{d} & 3$\sigma$ limit\tablefootmark{d} \\
\hline
12175     	       & $-$3.17 & $-$2.89 & 50$\pm$10 & $-1.67\pm0.16$ & $-$1.97 & $<-$2.70 ($-$1.97) & $<-$2.45 ($-$1.93) & $<-$2.23 ($-$1.88) \\
12729     	       & $-$2.35 &   \dots & \dots     & \dots  	& \dots   & $<-$1.65 ($-$0.55) & $<-$1.17 ($-$0.32) & $<-$0.70 ($-$0.01) \\
40789     	       & $-$2.88 & $-$3.06 & \dots     & \dots  	& \dots   & $<-$2.12 ($-$1.31) & $<-$1.75 ($-$1.23) & $<-$1.49 ($-$1.10) \\
40993     	       & $-$2.38 & $-$2.68 & \dots     & \dots  	& \dots   & $<-$2.12 ($-$1.33) & $<-$1.75 ($-$1.26) & $<-$1.51 ($-$1.14) \\
41460     	       & $-$3.10 & $-$2.78 & \dots     & \dots  	& \dots   & $<-$2.08 ($-$1.34) & $<-$1.70 ($-$1.23) & $<-$1.42 ($-$1.07) \\
41743     	       & $-$2.42 & $-$2.51 & \dots     & \dots  	& \dots   & $<-$2.14 ($-$1.53) & $<-$1.76 ($-$1.42) & $<-$1.50 ($-$1.27) \\
42096     	       & $-$2.60 & $-$2.40 & \dots     & \dots  	& \dots   & $<-$2.26 ($-$1.20) & $<-$1.87 ($-$1.14) & $<-$1.63 ($-$1.07) \\
42149     	       & $-$2.95 & $-$3.08 & 22$\pm$15 & $-2.03\pm0.42$ & $-$1.95 & $<-$2.49 ($-$1.70) & $<-$2.20 ($-$1.66) & $<-$1.99 ($-$1.61) \\
42241\tablefootmark{c} & $-$2.03 & $-$2.28 & 50$\pm$10 & $-1.69\pm0.15$ & $-$2.64 & $<-$2.68 ($-$2.12) & $<-$2.52 ($-$2.09) & $<-$2.37 ($-$2.04) \\
42324                  & $-$2.70 & $-$2.60 & \dots     & \dots       	& \dots   & $<-$1.88 ($-$0.88) & $<-$1.48 ($-$0.76) & $<-$1.18 ($-$0.58) \\
42795                  & $-$3.17 & $-$3.11 & \dots     & \dots       	& \dots   & $<-$2.24 ($-$1.38) & $<-$1.89 ($-$1.32) & $<-$1.64 ($-$1.23) \\
\hline                                    
\end{tabular}
\tablefoot{
\tablefoottext{a}{IDs,  [Fe/H], and [Ca/H] from Ad\'en et al. (2009a, 2011). 
}
\tablefoottext{b}{Corrected for the Fe blend (Sect.~3.3).} 
\tablefoottext{c}{This is Her-2 (K08).}
\tablefoottext{d}{Values in parentheses give the limit derived when accounting for an error in continuum placement (Sect.~3.5.1).}
}
\end{table*}
\subsection{Upper limits on equivalent widths for low $S/N$ spectra}
\label{sect:upperlim}
For the remaining spectra, no Ba\,{\sc ii} lines could be detected
 above the spectral noise.
 For these stars we chose to determine upper limits on the EWs and used them to derive upper limits on the
Ba abundances.
To this end, we followed the formalism of Cayrel (1988) to obtain a 1$\sigma$-error on the EW. This error can be propagated to an $n\sigma$ detection limit  for log\,$\varepsilon$(Ba).  
In practice, $\sigma$EW is a
function of the r.m.s. variation of the continuum (measured in
immediately adjacent continuum regions), the (Gaussian) full width at
half maximum (FWHM) of the line, and the pixel scale. 
The FWHM was estimated as 382 m\AA~from an S/N-weighted mean FWHM of other, stronger lines near the Ba~II~6141 \AA~feature, and the  Ba line itself in the three brighter targets. 
Table~1 lists these upper limits\footnote{The quoted values include a downward correction to account for the blend with a nearby Fe line; see Sect.~3.3.} 
for the 1, 2, and 3$\sigma$ detectabilities. 
Conversely, this procedure allows us to assess the significance of the actual line detections in the brighter stars as 
7.6$\sigma$ (\#12175),  2.8$\sigma$ (\#42149), and  
10.2$\sigma$ (\#42241). 
\begin{figure}[htb]
\begin{center}
\includegraphics[angle=0,width=1\hsize]{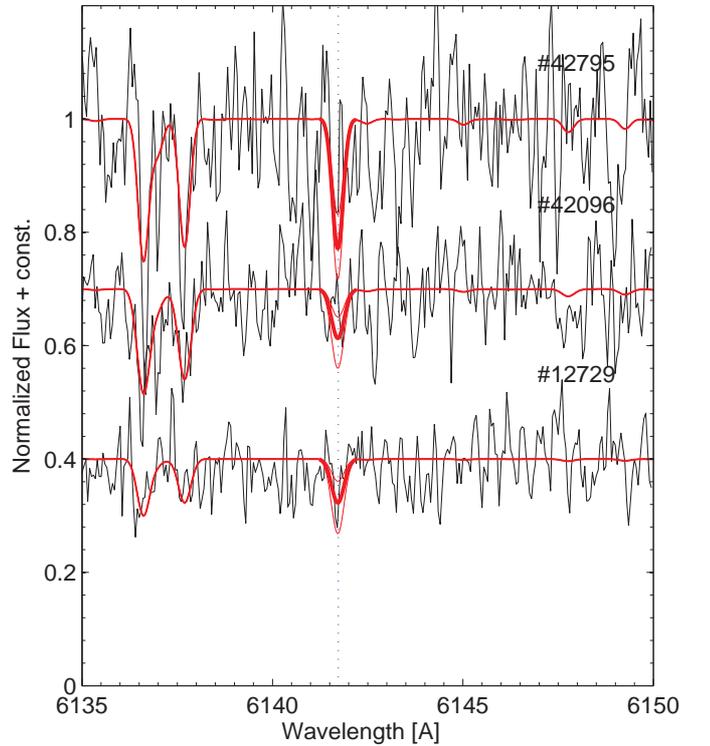}
\end{center}
\caption{Same as Fig.~1, but for three fainter stars (in order of
  descending metallicity). For these spectra we could only derive
  upper limits on log\,$\varepsilon_{\odot}$(Ba). The thick solid red
  lines  refer to the 3$\sigma$ limit of Table~1. The thin red lines
  shows the change when log\,$\varepsilon$(Ba) is varied by
  $\pm$0.4\,dex.}
\end{figure}
\subsection{A blending Fe\,{\sc i} line at 6141.730 \AA}

The Ba\,{\sc ii} line at 6141.713 \AA~is blended with an Fe\,{\sc i}
line at 6141.730 \AA. This needs to be accounted for when determining
log\,$\varepsilon$(Ba).  To this end, we computed synthetic spectra
for each star, using its derived stellar parameters and measured
[Fe/H] abundance. One synthesis was computed using only the Ba\,{\sc ii} line
and the second synthesis also included the Fe\,{\sc i} line.
The difference in the EWs measured from these two synthetic spectra
yields an indication of the level on contamination by the Fe\,{\sc i}
line (e.g., Dobrovolskas et al. 2012).
On average, we found a difference of 7\% in measured EWs,  with a maximum of 26\%~ for the most metal-rich star, \#42441 (Her-2). 
This leads to  [Ba/Fe] limits (3$\sigma$)
being lower by 0.03 dex on average. For  \#42441 the effect is stronger,
0.15\,dex. As \#42441 is the most metal-rich star in our sample and 
the strength of the Fe\,{\sc i} line is driven by the metallicity, this 
means that the effect is weaker in all other stars. 
This is also verified by the spectral syntheses in Sect.~3.4.

The situation is similar for the three detections and EW measurements of the blended line: 
For stars \#12175, \#42149, and \#42241 the effect on [Ba/Fe] is 0.02,  0.05, and 0.21 dex respectively 
and therefore comparable to the measurement uncertainties on our EWs (Table~1).
Thus, for the majority of the stars, in- or excluding the Fe blend does not significantly alter the values for the detection limit, with the possible exception of the most 
metal-rich target (\#42241), which we address in more detail in the next section. Moreover, the resulting decrease 
 would rather strengthen our conclusion of a very low Ba-level in this galaxy (Sect.~4). 
\subsubsection{Influence of errors on the Fe-abundance}
An error on the Fe abundance we assume for the model atmospheres will also affect the influence that the Fe-blend will exert on our  
estimate of log\,$\varepsilon$(Ba). 
Lowering [Fe/H] will diminish the blending Fe-line, resulting in an increase in the derived Ba abundance, and vice versa.
To assess this source of uncertainty, we quantitatively propagated an Fe-error by re-running the above procedures for a grid of Fe-abundances, 
covering $\pm$0.4 dex around the fiducial values. 

\vspace{2ex}\noindent
{\em \#42241:}
For the most metal-rich star the contamination will be strongest and this exercise can 
thus provide an upper limit to the variation of the derived  log\,$\varepsilon$(Ba) with [Fe/H].  
As a result, an Fe abundance lower by 0.3 dex translates into Ba-abundances 
higher by 0.27 dex, yielding an underestimate of [Ba/Fe] by 0.57 dex. Conversely, an increase in Fe by 0.3 dex means that the measured EW of the feature will be 
entirely due to iron with only negligible contribution from Ba; an error in Fe of 0.2 lowers log\,$\varepsilon$(Ba) by 0.34 dex. 

A11 estimated statistical errors on [Fe/H] of typically $\pm$0.15 dex, which is in line with the excellent agreement of 
  stellar parameters, metallicities, and/or iron abundances between
  different studies, based on various indicators\footnote{Amongst these measurements are [Fe/H]$_{\rm
    Stromgren}$=$-1.93\pm0.39$ (Ad\'en et al. 2009a); [Fe/H]$_{\rm
    CaT}$=$-2.39\pm0.13$ (K08); [Fe/H]$_{\rm I}$=$-2.02\pm0.04$ (K08);
  [Fe/H]$_{\rm I}$=$-2.03\pm0.14$ (A11), yielding an error-weighted
  mean of $-2.05\pm0.04$ dex.  }, 
  which strengthens the reliability of our measurement
 techniques. 
We can translate this into an error on log\,$\varepsilon$(Ba) of $^{-0.30}_{+0.18}$ dex and on [Ba/Fe] of $^{-0.45}_{+0.33}$ dex.

\vspace{2ex}\noindent
{\em \#12175:}
In addition, we tested the influence of varying Fe for the most metal-poor star, to which the Fe-blend contributes only  negligibly. 
As expected, even such large uncertainties on Fe of 0.4 dex do not change the measured  log\,$\varepsilon$(Ba) by more than 0.02 dex. 

We conclude that an error in Fe only affects the measurements at the metal-rich tail of our sample distribution, without notable influence on 
our claim of a low, overall Ba-abundance. 
In the figures below that show our results, we  indicate the error ellipse for \#42241 to illustrate this order-of-magnitude effect; as shown above, 
the error ellipses  on the more metal-poor stars based on  this source of uncertainty are negligible and are not shown.  
\subsection{Ba abundances from spectral synthesis}
We successfully determined Ba abundances using spectral synthesis for
the three brightest stars. For the remainder of the stars only upper
limits could be derived. The synthesis takes the blending Fe\,{\sc i}
line into account (see Sect.~3.3).

\paragraph{The three brightest stars:} 
The Ba\,{\sc ii} line in the two more metal-poor stars (of the three 
brightest stars), \#12175 and \#42149,
is well matched by the syntheses and yielded best-fit (in a least-squares sense) abundances
that differ from the EW measurement by
$-0.30$ and 0.08\,dex, respectively. As the $\chi^2$-procedures
indicate, a typical uncertainty of 0.4\,dex 
on [Ba/Fe]
can be reached in the fits
so that the above difference between EW-based values and the
synthesized spectra are well within the measurement errors (see also
Table~1). The best-fit spectra are shown in Fig.~1 
and the resulting Ba-abundance is listed in the column labeled ``Synth'' in Table~1.
In particular, the Fe lines between 6136 and 6138 \AA~are well reproduced.

The synthesis of the spectrum from star \#42241 (Her-2), however,
yields a very low [Ba/Fe] = $-2.78$. This lies well below the value
implied by the EW measurement (it is one dex lower). This cannot be
fully explained
 by the Fe-blend because the correction for the blend is only
0.21\,dex (Sect.~3.3), nor by the statistical measurement
uncertainty on the EW, which can only account for 0.15\,dex.
However, as the synthetic spectrum and the synthesis with
log\,$\varepsilon_{\odot}$(Ba)$\pm$0.4 dex of this object show, a
change in Ba abundance has hardly any influence on the depth and shape
of the line. Thus it appears that the blending Fe\,{\sc i} is the
dominant source for the observed absorption feature.  As a result, the
$\chi^2$ distribution saturates toward low Ba-abundances.
This was also confirmed by our test in Sect.~3.3.1.

\paragraph{Fainter stars:}
We also overplotted synthetic spectra on the noisy spectra for the
fainter stars, from which only upper limits could be gleaned, 
to judge the reliability of our method to derive the detection
limits. Three examples are shown in Fig.~2. As above, the Fe lines
blueward of the Ba\,{\sc ii} line are satisfactorily matched, while
the synthesis of the Ba\,{\sc ii} line itself indicates that our
3$\sigma$ limits are in fact good measures of the upper detection
limit in these spectra.
\subsection{Systematics}
\subsubsection{Continuum placement}
Originally, 
Cayrel's (1988) formula was devised for high-S/N spectra 
under the assumption that instrumental effects were corrected for, that the signal was photon-noise limited, and neglecting 
errors due to inaccurate sky subtraction, all of which are likely to be optimistic for our present, low-S/N spectra\footnote{In fact, the title of the respective proceedings 
volume was ``The Impact of Very High S/N Spectroscopy on Stellar Physics'' (Cayrel de Strobel \& Spite 1988).}). 
In fact, Cayrel states errors in continuum placement of 0.5\%, while our spectra are at a level of 3--8\% . 

When measuring EWs, we already accounted for this by allowing for a generous range in EW, estimated by eye, of 20\% and even 68\% for the weakest of the three detections (Table~1). 
For the upper limits from the noisier spectra we again followed Cayrel (1988) and calculated an error on the expected EW. The respective uncertainty in the placement 
was determined from the immediate continuum adjacent to the Ba-line. 
We then added the resulting limit in quadrature to the 
previously determined n$\sigma$-detection limits that only accounted for the noise characteristics of the line (see also Bohlin et al. 1983). 
These revised upper limits are listed (in parentheses) in Table~1.

Naturally, these limits are higher than those derived without accounting for the continuum placement; for instance, the difference amounts to 
0.44 dex on average for the case of 3$\sigma$-detections. 
As the continuum region around the  Ba line is essentially free of other blends in metal-poor stars (except for the Fe line discussed above), 
any such errors would in fact be mainly due to the (spline1-) normalization employed by A11. However, since we handled the region around the Ba line with 
care, we consider the upper limits that additionally include the continuum error as conservative high estimates. 
\subsubsection{NLTE, ionization, and level of significance}
We did not correct our results for departures
from LTE: The computations of Andrievsky et al. (2009) indicate that all
parameter changes ($\varepsilon$(Ba), [Fe/H], log\,$g$) conspire 
 to yield NLTE corrections no larger than 0.1 dex. The Ba\,{\sc ii} line at 
6141\,{\AA} 
used in the present work was, however, not included in their
calculations.  Following Short \& Hauschildt (2006), we estimated the 
downward NLTE corrections for the unsaturated Ba~II line at 6141 \AA,
at [Ba/H]$\sim -4$ as observed in our stars (Table~1), to be on the order of
0.06 dex, which even lowers our upper limits.  Likewise, Dobrovolskas et
al. (2012) found the same order of magnitude effect, albeit at higher
metallicities ([Fe/H]=$-$1.6 dex).

To account for the same dependencies on ionization
equilibrium and to remove sensitivities to the electron densities $N_e$ 
of the atmospheres (Seect.~3.6.2), one would strictly need to
normalize the abundance ratio of Ba~II relative to the iron abundance
from ionized species, i.e., [Ba~II/Fe~II]. However, the spectral range
of HR13 does not contain any Fe~II lines that are strong enough to be
reliably measured in the metal-poor red giant stars in the Hercules
dSph galaxy.  Accordingly, A11 did not list any [Fe~II/H] ratios and we
solely quote [Ba~II/Fe~I] for the three stars with actual
measurements rather than upper limits.  K08 derived an average
ionization imbalance of
log\,$\varepsilon$(Fe~I)$-$log\,$\varepsilon$(Fe~II)=$-0.22$ dex.
If this is representative of the entire Hercules RGB sample, our [Ba/Fe] abundance ratios should be shifted upward by this constant amount, leaving the conclusions of a  low Ba-level unaffected.  

An important concern for this study is the choice of $n$ -- the level
of significance in the detection limits.  Therefore, we computed
log\,$\varepsilon$(Ba) for $n=1, 2, 3$, which are listed in Table
1. Fig.~3 shows the element ratios of [Ba/H] and [Ba/Fe] for the cases
of 2 and 3$\sigma$.  In the following we continue by adopting the
3$\sigma$ values as a statistically sound detection limit.
\subsubsection{Uncertainties on stellar parameters}
In computing upper limits we are dealing only with the noise
characteristics of the spectra and thus with random errors. While it
may seem unusual to quote systematic uncertainties on the statistical
detection limits, these can give us a deeper insight into the actual
level of depletion of chemical elements in the Hercules stars.  Therefore we performed a
standard error analysis in which we separately varied each stellar
parameter in the atmospheres by their typical uncertainties, as
estimated by A11, and re-computed the detection limits as before. In
this way, we can state an ``upper limit on the upper limits'', i.e.,
the highest ($n\,\sigma$) detection limit assuming that all 
stellar parameters were significantly in error.  We carried this out
for star \#42241, as an example of a measurement based on an actual EW
detection, and for \#42795, the most metal-poor star in the sample
(Table~1).  Table~2 lists these limits for the 2\,$\sigma$ and
3\,$\sigma$ cases upon varying the stellar parameters.
 \begin{table}[htb]
\caption{Detection limits upon systematic variation of the stellar parameters.}             
\centering          
\begin{tabular}{lccc}     
\hline\hline       
& \#42241 & \multicolumn{2}{c}{\#42795} \\  
\cline{3-4}
 \raisebox{1.5ex}[-1.5ex]{Parameter}  & log\,$\varepsilon$(Ba) (EW) & log\,$\varepsilon$(Ba) (2$\sigma$) & log\,$\varepsilon$(Ba) (3$\sigma$) \\  
\hline
T$_{\rm eff}$  +  100 K             & $-$1.70 & $<-1.96$ & $<-1.71$ \\
T$_{\rm eff}$ $-$ 100 K             & $-$1.66 & $<-1.83$ & $<-1.58$ \\
log\,$g$       + 0.35  dex          & $-$1.83 & $<-1.99$ & $<-1.75$ \\
log\,$g$      $-$ 0.35 dex          & $-$1.56 & $<-1.79$ & $<-1.54$ \\
$\xi$          +  0.2 km\,s$^{-1}$  & $-$1.66 & $<-1.88$ & $<-1.63$ \\
$\xi$         $-$ 0.2 km\,s$^{-1}$  & $-$1.72 & $<-1.90$ & $<-1.66$ \\
$[$M/H$]$          + 0.2  dex       & $-$1.75 & $<-1.90$ & $<-1.65$ \\
$[$M/H$]$         $-$ 0.2 dex       & $-$1.63 & $<-1.88$ & $<-1.64$ \\
$[\alpha$/Fe$]$ = 0                 & $-$1.78 & $<-1.89$ & $<-1.64$ \\
\hline       	
\end{tabular}	
\end{table}	

Changes in gravity lead to the
largest uncertainties for star \#42241, with $\Delta$\,log\,$\varepsilon$(Ba) = $\pm$0.13
dex. Summing all contributions in quadrature, we obtain a total
systematic uncertainty on the Ba abundance of this star of 0.16 dex,
which is comparable to the random error based on the EW measurement
uncertainty.

For star \#42975, a comparison of Table~2 with the default results in
Table~1 shows that, again, the surface gravity plays the dominant role in
the Ba abundance determinations. The combined systematic uncertainty
in this case amounts to 0.13 dex for both 2$\sigma$ and 3$\sigma$.  We
can rephrase this such that, only if all the errors on the stellar
parameters were to conspire to raise the values for
log\,$\varepsilon$(Ba), we would obtain upper limits that were 0.13
dex higher at the maximum. This does not affect our conclusions of an
overall low Ba level in the Hercules dSph galaxy.
\subsubsection{Opacity distribution functions}
For either level of significance considered above, switching between $\alpha$-enhanced and solar-scaled opacity distributions does not have a noticeable effect on the Ba abundances or the limits thereof (see also Sect. 3.6.2). 
Thus there is no apparent, systematic effect on the abundance limits due to the large spread in the $\alpha$-elements in Hercules, as represented by Ca (A11). 
		
Koch et al. (2008) found that their two stars had highly elevated [Mg/Fe] and [O/Fe] ratios of up to 0.8 and 1.1 dex, which lie significantly higher than the standard $\alpha$-enhancement of 0.4 dex assumed in this work. 
However, at the overall low iron abundances of the Hercules stars below $-2$ dex,  the dominant source of electrons is the ionization of hydrogen, which 
certainly holds for the most metal-poor stars  at  [Fe/H] $\sim -3$ dex (cf. Koch \& McWilliam 2008). 

Secondly, in the most metal-rich target of the sample, we find a slope of $\Delta$\,log\,$\varepsilon$(Ba) / $\Delta\,[\alpha$/Fe] of 0.23 (Table~2), which means that extrapolating the transition from solar to $\alpha$-enhanced models 
to the observed  extreme O/Fe abundance of this star, \#42241, would yield a Ba abundance higher by 0.16 dex.   Similarly, for our test on the most metal-poor sample star we obtain a negligible increase in 
log\,$\varepsilon$(Ba) of 0.02 dex, provided it had the same strong level of $\alpha$-enhancement as the two stars measured by K08.
\section{Results}
Figure~3 shows our results (after accounting for the Fe-blend). We show the Ba abundances obtained 
from the EW measurements and 
upper detection limits  for the 2 and 3$\sigma$ cases. We also
include, for comparison, Galactic stars from the compilation of Venn
et al. (2004; and references therein) and Milky Way dSph satellites
including the more luminous dSphs (Shetrone et al. 2001, 2003, 2009;
Fulbright et al. 2004; Sadakane et al. 2004; Geisler et al. 2005; Aoki
et al. 2009; Cohen \& Huang 2009; Frebel et al. 2010a; Letarte et
al. 2010; Tafelmeyer et al. 2010; Lemasle et al. 2012; Venn et
al. 2012; and {Starkenburg et al. 2012}) and the most recent measurements in a few ultrafaint dSphs
(K08; Feltzing et al. 2009; Frebel et al. 2010b; Norris et al. 2010b;
Simon et al. 2010; and {Gilmore et al. 2013}).
\begin{figure*}[htb]
\begin{center}
\includegraphics[angle=0,width=1\hsize]{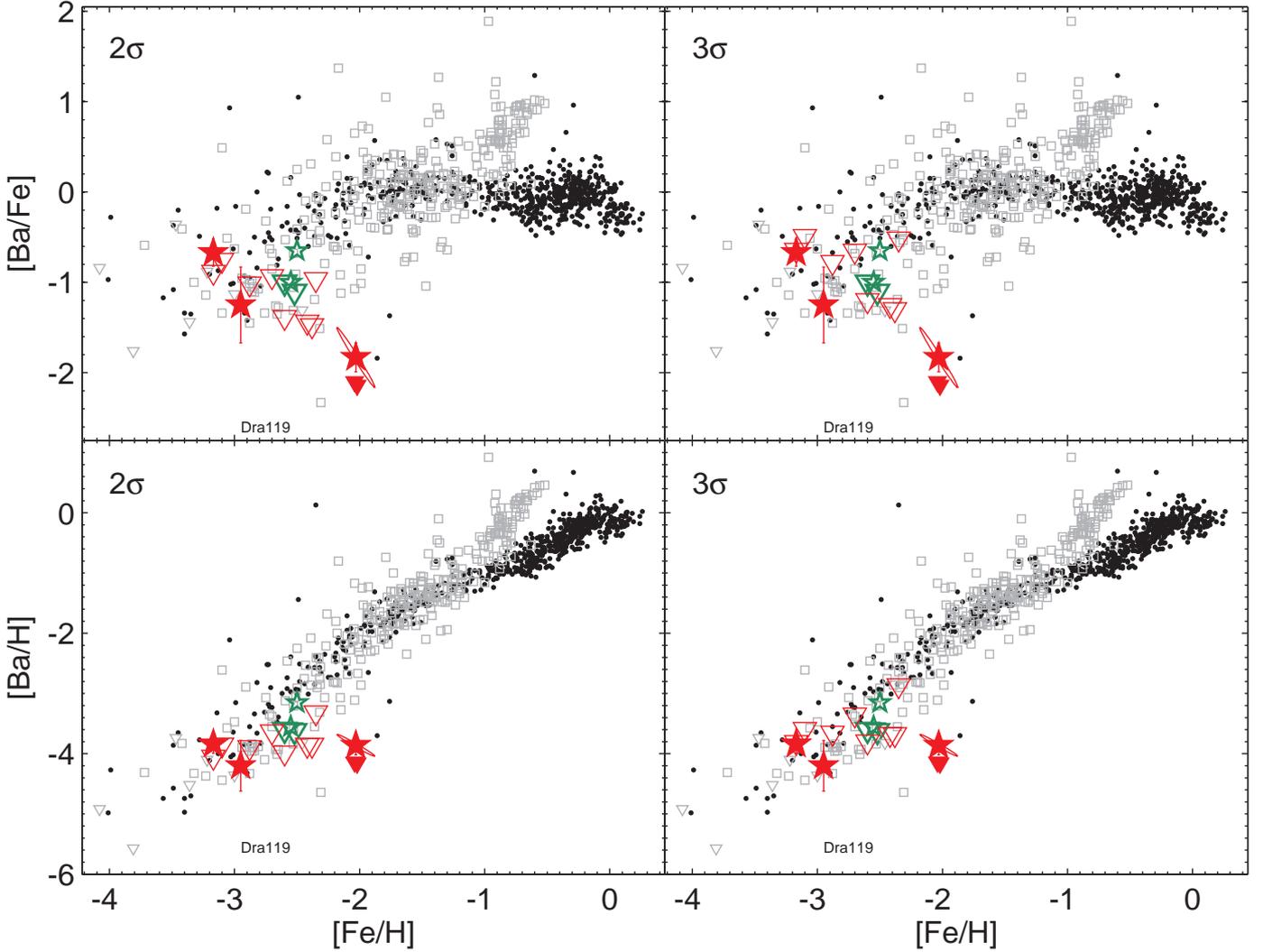}
\end{center}
\caption{Barium abundance results from the present study. Red filled star symbols show actual detections and EW measurements in Hercules (including total measurement uncertainties), while open triangles denote upper limits. 
The left (right) panel shows 
2$\sigma$ (3$\sigma$) detections. 
Red ellipses on the most metal-rich target indicate the error ellipses due to the Fe-blend; this source of contamination is much smaller for the more metal-poor stars (Sect.~3.3.1). 
The red solid triangles depict the two stars from K08. 
Gray squares are data for dSphs from the literature and the black points depict Galactic halo and disk stars from the
compilation of Venn et al.  (2004). Green symbols refer to the detections (stars) and upper limits (triangles) 
in Hercules
by Fran\c cois et al. (2012).  We also mark, by text, the location of the  star Dra 119 that is deficient in  heavy-elements (Fulbright et al. 2004).
}
\end{figure*}

The Ba abundances and limits are very low, at a mean [Ba/H] of $-$3.97
from the three EW measurements, and $<$[Ba/H]$>$ $< -3.84$
(2$\sigma$), $< -3.55$ (3$\sigma$), respectively. 
Thus, the Hercules dSph galaxy appears strongly depleted in Ba and, including
the non-detection of Sr and other generally strong features in the
red giant spectra of K08, we infer that Hercules is overall deficient in
neutron-capture elements.  The values measured here overlap with
metal-poor halo stars at similar metallicities and about three dozens
of dSph and UFD stars, but keeping in mind that our values are mostly
upper limits, it is likely that the Hercules UFD shows an even stronger
depletion than these metal-poor environments.

Intriguingly, the low Ba abundances persist over the full range of
metallicity  in the Hercules dSph galaxy.
Therefore, Ba is essentially (almost constantly) low and the only
reason for the decline in [Ba/Fe] towards higher [Fe/H], seen in the
top panels of Fig.~3, is the onset of the Fe contributions, either
from low-level Fe production in SNe\,Ia or in smaller
amounts from massive stars, but coupled with an inhomogeneous
mixing. In fact, the broad spread in iron (with a full range in
[Fe/H] of 1.14 dex; A11) is typical of the low luminosity, dark-matter dominated dSphs (Koch 2009; Ad\'en et al. 2009; Koch et
al. 2012a).  Taken at face value, the peak-to-peak variation in [Ba/H]
across the full range in Fe is a mere 0.94 dex
(3$\sigma$) 
-- hardly an enrichment worth mentioning.

For a fair comparison, we reproduce the plots in Fig.~4, where the upper limits from our spectra (red triangles) are those after including a conservative error on the 
continuum placement.   
\begin{figure}[htb]
\begin{center}
\includegraphics[angle=0,width=1\hsize]{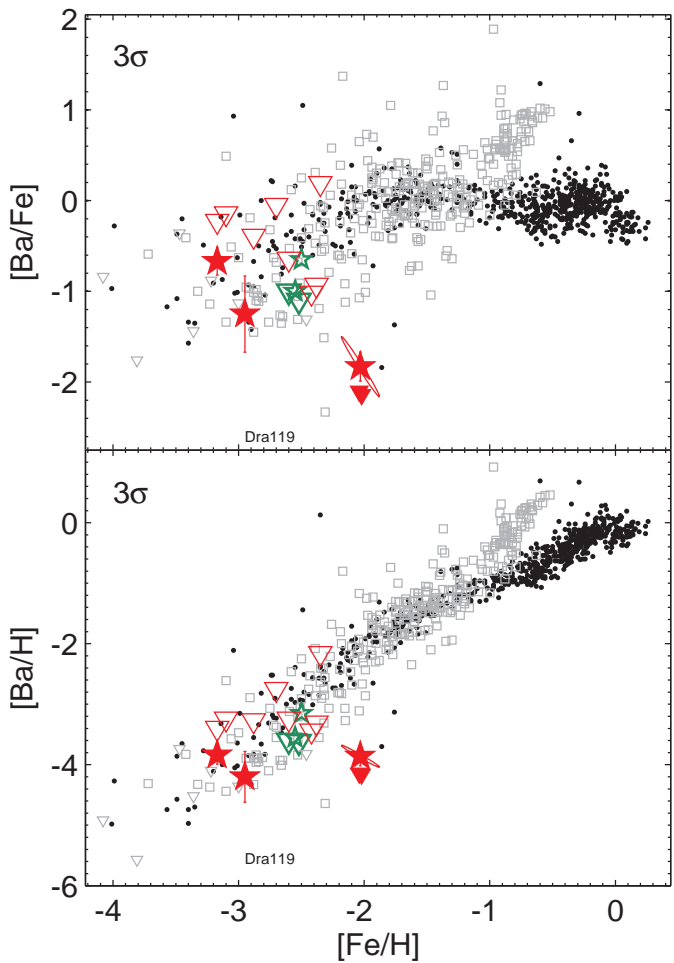}
\end{center}
\caption{Same as Fig.~3 for the 3$\sigma$ detections. Here, upper limits include a conservative error on the continuum placement (Sect.~3.5.1).}
\end{figure}
An increase in the upper limits due to an additional, large account of continuum placement errors would bring the upper limits
in better accord with metal-poor halo and dSph stars (Fig.~4). However, the most metal-rich star, \#42241 / Her-2,  remains 
a puzzle in that it is strongly depleted in Ba, no matter how conservatively we construct its error budget. Its strong deficiency 
in [Ba/H] drives the impression of the overall, low Ba-abundance we see in this galaxy. 
\section{Comparison with the literature}
Table~3 provides a comparison between our  results and results from the literature. 
\begin{table*}
\caption{Comparison with the literature}             
\centering          
\begin{tabular}{cccccccccccc}     
\hline\hline       
 & \multicolumn{2}{c}{\#42241} && \multicolumn{2}{c}{\#41460} && \multicolumn{2}{c}{\#42149} && \multicolumn{2}{c}{\#42795} \\
\cline{2-3}\cline{5-6}\cline{8-9}\cline{11-12}
              & K08      & A11/TW    &&     A11/TW &	  F12  &&    A11/TW &	  F12 &&    A11/TW &       F12 \\
\hline
ID            & Her-2    &     42241 &&      41460 &	466011 &&     42149 &  400664 &&     42795 &	400508 \\
$[$Fe I/H$]$  & $-$2.02  &   $-$2.03 &&    $-$3.10 &   $-$2.55 &&   $-$2.95 & $-$2.50 &&   $-$3.17 &   $-$2.60 \\
$[$Fe II/H$]$ & $-$1.78  &     \dots &&      \dots &	 \dots &&     \dots &	\dots &&     \dots &	 \dots \\
$[$Ca/H$]$    &	$-$3.15      &   $-$2.28 &&    $-$2.78 &	 \dots &&   $-$3.08 & $-$2.35 &&   $-$3.11 &   $-$2.45 \\
$[$Ca/Fe$]$   & $-$0.13	 &   $-$0.25 &&       \phs0.32 &	 \dots &&   $-$0.13 &	 \phs0.15 &&      \phs0.06 &	  \phs0.15 \\
$[$Ba/H$]$\tablefootmark{a}    & $<-2.14$ & $< -4.51$ &&  $< -3.59$ & $< -3.55$ && $< -4.16$ & $-$3.15 && $< -3.81$ & $< -3.60$ \\
$[$Ba/Fe$]$\tablefootmark{b}   & $<-3.92$ & $< -2.51$ &&  $<- 0.49$ & $< -1.00$ && $< -1.21$ & $-$0.65 && $< -0.64$ & $< -1.00$ \\
\hline       	
\end{tabular}
\tablefoot{
\tablefoottext{a}{Values from this work are quoted as 3$\sigma$-limits.} 
\tablefoottext{b}{Relative to Fe~I, except for the values from K08.} 
\tablebib{K08:~Koch et al. (2008a); A11 (for IDs, Fe, and Ca):~Ad\'en et al. (2011); F12:~Fran\c{c}ois et al. (2012); TW (for Ba):~This Work.}
}
\end{table*}	

At first glance it may appear counterintuitive that the actual EW
detections in the present work lie above the {\em upper} limits placed
by K08 (shown as red solid triangles in Fig.~3).  However, one should
bear in mind that the spectral range (at a resolving power similar to
the one of the FLAMES spectra in this work) of K08 also permitted the
use of the Ba~II 6496.91 \AA~line, which is inaccessible on the
limited spectral coverage of HR13. Furthermore, K08 adopted a
2$\sigma$ detection limit to state their Ba-abundance results.

Fran\c{c}ois et al. (2012) reported on detections of Ba in two red
giants in the Hercules dSph galaxy, yielding [Ba/Fe]$\sim -0.3$ dex, and also upper limits
on the same order of magnitude for two other stars.  Three of their
stars overlap with our sample. Their Ca abundances agree very
well with the data of A11 to within the errors.  However, the iron
abundances by Fran\c{c}ois et al. (2012) are higher by 0.5 dex on
average then A11's measurements.
While these authors remarked that they ``did not find extreme abundances
in our Hercules stars as the one found by Koch et al. (2008)'', their
values for [Ba/H] 
agree with the present study. Thus the [Ba/H]
abundances we measure are lower by 0.42$\pm$0.30 dex than the values
shown by Fran\c{c}ois et al. (2012).
Reasons for any discrepancies can be sought in the different
instruments and, hence, resolutions of both studies. In particular,
the VLT/X-shooter (Vernet et al. 2011) spectra of Fran\c{c}ois et
al. (2012) have lower resolutions ($R\sim$8000--12000) than the
present FLAMES data ($R=20000$).  Moreover, it may not be a
statistically sound measure to compare upper limits from different
sources in a one-to-one fashion.
\section{Discussion}
Unfortunately, our knowledge of the chemical fingerprints of the
enrichment from a few or even single SNe are based on very few stars
in dSphs and UFDs  
(only two stars in Hercules, three in Leo~IV and Boo~I, and the extreme Dra~119 
in the Draco dSph), and a handful of
extremely metal-poor halo stars with [Fe/H] around $-$5 to $-3$ dex
(e.g., Aoki et al. 2006; Cohen et al. 2008).
In the present paper we aim to expand our knowledge on the enrichment
patterns and chemical enrichment histories of dSph galaxies by providing
 more data on the neutron-capture element Ba in the Hercules UFD. 
\subsection{Enrichment patterns from SN in dSph galaxies}
As the number of detailed abundance analyses of dSph galaxies studied using high-resolution
spectra has grown, the number of stars with ``anomalous'' abundance patterns has also
increased. Fulbright et al. (2004) found the first such star --
Dra~119 (indicated in Fig.~3). Also this star is strongly depleted in $s-$process elements. More
recently, another star with unusual
abundance patterns, S1020549, was found in the Sculptor dSph galaxy (Frebel et al. 2010). Also this
star is low in $s-$process elements. Other studies have found unusual
[Mg/Ca] ratios in the UFDs, e.g., in Boo~I (Feltzing et al. 2009).

Typically, the interpretation of the observed abundance patterns in
these few unusual and metal-poor stars are matched to the predictions
of theoretical yields to estimate how many SNe were involved in
the enrichment of the Interstellar Medium (ISM) that that particular star or stars formed
from.  Figure~3 shows the Ba abundances for many stars, including
Dra~119. This star is peculiar in that it is depleted in Sr and Ba by
at least two decades (Fulbright et al. 2004). The majority of other
heavy elements (Z$\ge$30) are systematically lower by three to four times
compared with halo stars of similar metallicity.
Since the light
element patterns in Dra~119 are consistent with enrichment from a
massive ($\sim$20 M$_{\odot}$) SN~II event the conclusion, perforce,
is that massive, metal-poor SNeII do not contribute to the $r$-process
in proto-dSphs/UFDs~\footnote{Conversely, the most likely mass range
  for SNe II to host the $r$-process lies around 8--10 M$_{\odot}$
  (e.g., Qian \& Wasserburg 2007).}. Thus even if the light elements
are consistent with a massive\footnote{A general concern in dealing with 
the the earliest chemical enrichment
histories and the ensuing nucleosynthesis by massive stars is the
question of ``how massive is massive?''.
High-mass SNe models (e.g., employing prompt enrichment and accounting
for pair-instability SNe) in the range of $>100$ M$_{\odot}$ failed to
reproduce many (light, Fe-peak, and neutron-capture) elements in
Dra~119 by more than an order of magnitude
and recent simulations in fact rather support ``ordinary massive''
first stars of a few tens M$_{\odot}$ (e.g., Hosokawa et al. 2011,
2012).
}, single progenitor,  the neutron-capture
elements appear to tell a different story.

S1020549 in Sculptor has a Ba abundance as low as that of Dra~119,
albeit its metallicity is lower than that of Dra~119 by one dex
(Frebel et al. 2010). It may appear tempting to compare these two
objects by considering Dra~119 to have experienced enrichment channels
similar to S1020549 with an additional, later contribution of
iron. However, the notably high Mg abundance in Dra~119 seems to
preclude this analogy.  Furthermore, the Scl dSph galaxy is more
luminous by $\sim$2.2 mag, thus providing a rather different
environment than the Draco dSph galaxy and, on the relevant, small
scales, invoking different chemical enrichment histories.

One star in Leo~IV is significantly underabundant in the $n$-capture
elements Ba and Sr, similar to what was found in other UFDs.  Simon et
al. (2010) argued that the chemical evolution of these environments was
universal and consistent with predictions for a population~III SN
explosion.

However, in none of the dSphs and UFDs discussed above was it found
that these low levels of, e.g., Ba were traced across their full
metallicity ranges. However, this is now the case for
the Hercules UFD, where we see low Ba across the entire broad range of
metallicities (Fig.~3).  As indicated above, the Hercules stars share a
common locus with some 36 stars in the low-metallicity
([Fe/H]$\lesssim -2.2$ dex), low-Ba ([Ba/H]$\lesssim -2.6$ dex)
regime.  Out of these, 25 are distributed over six of the more
luminous dSphs, while 11 stars found in four of the UFDs (Boo~I, Com,
Leo~IV, and UMa~II) show similar levels of depletions.  In particular, 
one star in the Com UFD is similarly depleted in Ba as \#42241
(Her-2), at [Fe/H] = $-2.3$.  Thus all the UFDs with heavy element
abundances published to date host at least one and often a few stars
with a significant lack of the $n$-capture elements.

Less is known about seriously depleted stars in the Milky Way. One
example is provided by the mildly metal-poor halo star BD+80$^{\circ}$
245 (Carney et al. 1997),  which resembles the most metal-rich Hercules star,
\#42241, in the [Ba/H]-vs-[Fe/H] plane.  While it shows [Ba/Fe] (and
[Eu/Fe]) ratios lower by 1 dex than other halo stars at similar metallicities, its significantly subsolar $\alpha$-element abundances
render it not fully comparable with the stars discussed above, all of  which
appear to have normal levels of the lighter elements.  This
perhaps implies that different enrichment mechanisms might be at play
(Ivans et al. 2003), possibly the mixing of SNe\,II ejecta with
contributions from some SNe\,Ia material.
\subsection{Hercules' chemical enrichment history}
As discussed above, strong depletion of heavy elements is also found in
individual stars of other UFDs. The current study suggests for the
first time, however, that a very low Ba abundance is found {\em within an
  entire} dSph, systematically over a broad range in metallicity.
How can we then account for a spread in iron whilst there is no
accompanying spread in the heavy element barium?

The metallicity distribution function (MDF) for the Hercules dSph
galaxy is truly metal poor. The most metal-rich stars have [Fe/H] of
$-2$ dex and extend down below $-$3 dex (Ad\'en et al. 2009, 2011, 2013 
[submitted]). This makes it perhaps one of {\it the} most metal-poor systems
known and thus an interesting laboratory for testing our understanding
of galactic chemical evolution.  In fact, based on the overall low
metallicities, A11 suggested that star formation in Hercules ceased early
after the galaxy formed.

Moreover, there are reliable determinations of [Ca/Fe] for some 15
stars in the Hercules dSph galaxy (K08; A11; Fran\c{c}ois et al. 2012).
These show the typical trend with a flat part at the lowest [Fe/H]
followed by a knee and an ``early'' (i.e., relatively metal-poor)
downturn toward increasing [Fe/H] (e.g., Fig.~10 in A11; see also
Fig.~5).  The canonical interpretation is that of a somewhat extended
star formation history with contribution from SNe\,II as well as from
SNe\,Ia. However, the star formation rate must have been weak for
SNe\,Ia to be able to contribute at such low metallicities.  In this
context, Koch et al. (2012b) argued that chemical evolution models
(similar to those of Lanfranchi \& Matteucci 2004) with a low star
formation efficiency around 0.01 -- 0.001 Gyr$^{-1}$ match the
observed trends in [Ca/Fe] reasonably well (see also Fig.~5).
\begin{figure}[htb]
\begin{center}
\includegraphics[angle=0,width=1\hsize]{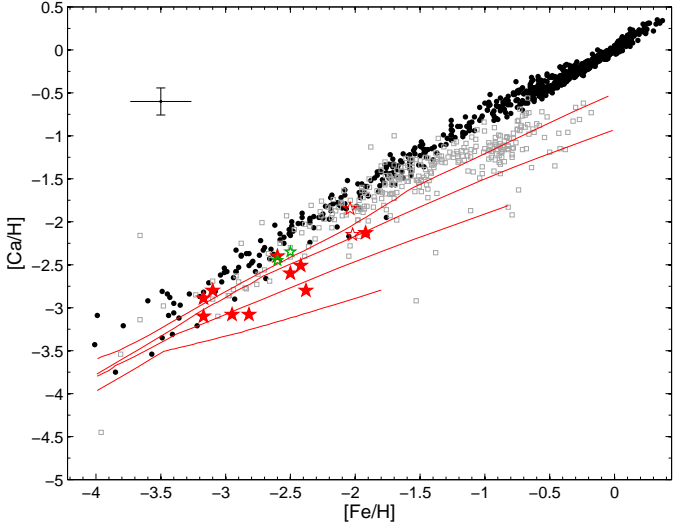}
\end{center}
\caption{Ca abundances from the same sources as in Fig.~3, with Fe and
  Ca values for the Hercules UFD taken from A11. The red lines are chemical
  evolution models for Hercules (Lanfranchi \& Matteucci 2004; Koch et
  al. 2012b), assuming star-forming efficiencies $\nu$ of (top to
  bottom) 0.1, 0.01, 0.001, and 0.0001 Gyr$^{-1}$.}
\end{figure}

Based on a comprehensive abundance analysis of two RGB stars in 
the Hercules dSph galaxy, K08 found that these stars possibly formed
from gas that had only been enriched by one or a few massive SNe
events (see also Simon et al. 2010).  This scenario can now be tested against the broad range in both
Fe as well as in Ca -- can a single SN account for the entire Fe, Ca, and
Ba we measure in Her?

We adopted the low luminosity and the stellar mass derived for the
Hercules dSph galaxy by Martin et al. (2008). The stellar mass is very low at
4--7$\times10^{4}$ M$_{\odot}$. Additionally, we adopted the MDF from
A11. With this information its possible to derive that we can expect
Hercules to contain a few times 0.1 M$_{\odot}$ of Fe\footnote{The actual
  value depends on the chosen IMF of Martin et al. (2008) and the use
  of A11's spectroscopic or photometric MDF.  In practice, M(Fe)
  ranges from 0.076--0.28 M$_{\odot}$.}.  Similarly, using the [Ca/H]
distribution of A11, Hercules contains a total of $\sim$0.005--0.01
M$_{\odot}$ in Ca. 
Drawing Ba abundances from our observed upper limits,
log\,$\varepsilon$(Ba)$< - 1.15$, translates into a total Ba-content
of at most  a few times $10^{-6}$ M$_{\odot}$.  Comparison with the
yields of, e.g., Heger \& Woosley (2010) shows that these amounts of
Fe and Ca can already be produced by 1--3 SNe~II of $\la$20
M$_{\odot}$, accompanied by little to no Ba.
This does not include the dilution of any such ejecta with the gas content of the early Hercules, however, 
and the accompanying question of the pre-enrichment of the early ISM\footnote{For instance, Tassis et al. (2012) 
argued that hypothesis that the UFDs were pre-enriched by a first generation of stars implies that essentially {\em all} heavy elements observed in 
systems with M$_{*}\le10^5$ M$_{\odot}$ (such as Her) have been produced by the Pop III stars.}  is beyond the scope of the present data and paper. 

In Fig.~6 we compare our measurements with the same chemical evolution models, following Lanfranchi \& Matteucci (2004), as introduced, for Ca, in Fig.~5 above. 
\begin{figure}[htb]
\begin{center}
\includegraphics[angle=0,width=1\hsize]{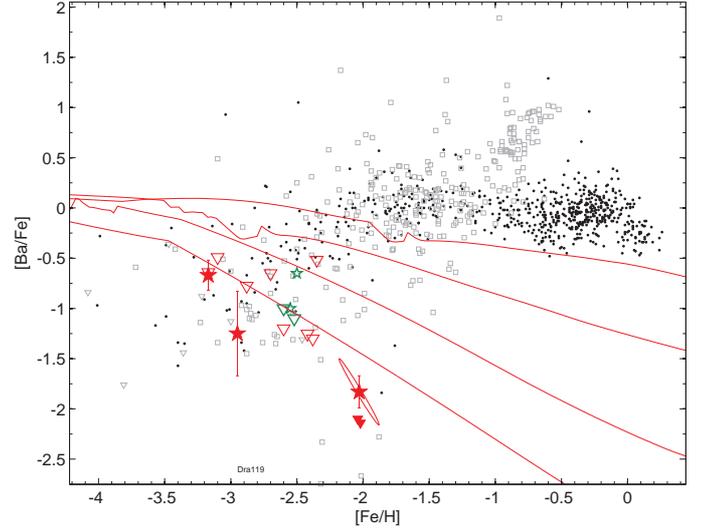}
\end{center}
\caption{Same as Fig. 3 (top right; 3$\sigma$). The red lines are based on  the same chemical
  evolution models as shown in Fig.~5  (Lanfranchi \& Matteucci 2004; Cescutti et al. 2006) with star-forming efficiencies $\nu$ of (top to
  bottom) 0.1, 0.01, 0.001, and 0.0001 Gyr$^{-1}$.}
\end{figure}
Contrary to the Galactic halo trend, the predicted [Ba/Fe] ratio in the Hercules UFD is decreasing with [Fe/H] -- as is indeed observed. 
The reason is that the models for Hercules assume an extremely low star formation rate and that star formation was extended in time.
Ba is produced in part from massive stars as an $r$-process element (Cescutti et al. 2006) and by the $s$-process in stars with masses between 1.5--3 M$_{\odot}$ with life times of 400 Myr and longer. 
In this scenario the massive stars in the Hercules dSph galaxy contribute at very low [Fe/H], very modestly, to both Ba  and Fe. 
When the Type Ia SNe start restoring the bulk of Fe, the Fe abundance in the ISM is still very low because of the low star formation efficiency and the [Ba/Fe] ratio start to decrease, as it happens in the Milky Way 
 at much higher metallicity. Thus the trend seen in Hercules appears as a natural consequence of the time-delay model coupled with a very low star forming rate. 
This conclusion remains valid even when using the high upper limits of Fig.~4, which resulted from large continuum uncertainties. 

However, our results might imply that the ``constant'' [Ba/H] in Fig.~3, or the
decline in [Ba/Fe] with increasing metallicity, respectively, could be solely
driven by the most metal-rich stars in the samples (Her-3,
Her-2/\#42441), which is found both in the results from K08 and in the
present study.
We cannot exclude the possibility that these ``outliers'' are too
metal-rich analogs of the remaining, more metal-poor sample, which
could have received an extra boost in Fe by a nearby massive SNe
event that had formed from pre-enriched gas.
\section{Summary}
We have obtained direct measurements and upper limits for the heavy element abundances of the Hercules UFD as traced by the $s$-process element barium. 
Similar to other faint dSphs and halo field stars of similar metallicity, Hercules shows low levels of Ba
 (see also Roederer 2013), 
but this depletion has now been traced over its full metallicity range. 

Without doubt, the Hercules dSph galaxy suffered from a low star-formation efficiency. 
Ultimate tests for chemical evolution involving sets of yields, an IMF, and Hercules' star formation history, 
accounting for enrichment by SNe II and SNe Ia and AGB (if any) are clearly warranted, but are inevitably limited by the currently sparse chemical abundance 
information at present. Measuring Mg in a large stellar sample and placing limits on more $n$-capture elements, preferably Sr 
(Roederer 2013), is crucial for a deeper insight into Hercules' enrichment history. 
Unfortunately, the spectral range of the HR13 spectra available to us for this study did not permit us to measure Mg so that we cannot judge 
whether the remainder of our sample  also has elevated [Mg/Ca] ratios, as found in  Her-2 and Her-3 of K08, which would be indicative of the occurrence of only a  few massive SNe events. 
Clearly, a definite reconstruction of the small-scale chemical enrichment history of this (and other) UFD needs to await the 
next generation of telescopes and spectrographs to obtain clear measurements of heavy elements above the  noise. 
Specifically, deeper studies of the Ba abundance itself would benefit from targeting other, intrinsically stronger Ba-lines at 5853 and 6494 \AA, and the strong 
resonance doublet lines at 4554 and 4934 \AA. 
\begin{acknowledgements}
AK acknowledges the Deutsche Forschungsgemeinschaft for funding from  Emmy-Noether grant  Ko 4161/1. 
We thank P. Fran\c{c}ois  for sharing his results before publication and a very helpful referee report. 
A. McWilliam, I.U. Roederer, H.-G. Ludwig, R.P. Church, T. Brown, and A. Frebel are thanked for helpful discussions. 
\end{acknowledgements}
\end{document}